\documentclass{jps-cp}
\usepackage{txfonts} 

\title{Quantification of the Flavor Diagonal Hadronic CP Violation}

\author{Nodoka \textsc{Yamanaka}$^{1,2}$}

\inst{$^{1}$Department of Physics, Graduate School of Science, Tohoku University, Sendai 980-8578, Japan \\
$^{2}$Nishina Center for Accelerator-Based Science, RIKEN, Wako 351-0198, Japan}

\email{nodoka.yamanaka@riken.jp}

\recdate{January 12, 2026}

\abst{
The flavor diagonal CP violation of elementary particle physics contributes to the atomic, nuclear, and nucleon electric dipole moments (EDMs), T-violating neutron optics, and to the angular correlations of beta decay. 
In this contribution, we review the basics and the importance of CP violation in the search for new physics beyond the standard model, the recent progress in the quantification of the hadron level CP violation contributing to the aforementioned observables, and finally the current attempt to solve the strong CP problem without additional interactions and fields.
}

\kword{CP violation, beyond the standard model, chiral perturbation theory, strong CP problem}

\begin{document}
\maketitle

\section{Introduction}

The CP transformation \cite{Landau:1957pbu} is the combination of the parity transformation (P), the simultaneous inversion of all spatial coordinates $\vec x \to -\vec x$, and the charge conjugation (C), which consists of interchanging particles and their antiparticles.
The violation of CP has been an important topic in particle physics since its discovery in the decay of $K$ mesons \cite{Christenson:1964fg}.
It is also one of the three necessary conditions for the realization of the matter abundance of the present Universe \cite{Sakharov:1967dj}.
Due to this scientific importance, there are nowadays many experimental approaches to search for new sources of CP violation beyond the standard model (SM), notably the measurements of the electric dipole moment (EDM) \cite{Graner:2016ses,Abel:2020gbr,Roussy:2022cmp,Yamanaka:2014mda,Yamanaka:2017mef,Chupp:2017rkp}, low energy T-violating neutron optics \cite{Nakabe:2025ehg,Gudkov:1991qg,Bowman:2014fca,Gudkov:2017vqn}, and beta decay \cite{Huber:2003gr,Soldner:2004xm,Mumm:2011nd,Kozela:2011mc,Chupp:2012ta,Murata:2016bzr,Herczeg:2001vk,Gonzalez-Alonso:2018omy}.
In this contribution, a short review of the theoretical aspects of CP violation at low energy is presented.
In the next section, we will introduce the principle of CP violation and its necessity in looking for new physics.
In Sec. \ref{sec:hadronicCPV}, we summarize the current status of the evaluation of hadron level CP violating interactions without flavor violation.
In Sec. \ref{sec:strongCP}, we review the resolution proposed for solving the strong CP problem within the SM.
The final section is devoted for the conclusion.

\section{CP violation}

In this section, we introduce the P, C, and CP symmetries, as well as their importance in the experimental search for new physics.
The operation of P to the Dirac fermions which include quarks and leptons is trivial: we just exchange the left- and right-handed chiralities, as $\psi_L \leftrightarrow \psi_R$.
Applying C to a Dirac field changes the sign of the complex phase as $\psi \leftrightarrow \psi^*$.
Here the subtle point is that the chirality, or the spin in nonrelativistic physics, which is in the classical interpretation the rotation of the charge of the particle around a given axis (suppose $z$), is a complex phase modulation $\psi_L \equiv e^{i(j_z \phi+ \delta)} ||\psi||$, where $j_z =\pm \frac{1}{2}$ is the $z$-component of the angular momentum of the fermion, and $\phi$ is the angle around the $z$-axis, and we assumed that $||\psi ||$ contains all other informations than the complex phase.
The C transformation then changes $\psi$ as 
\begin{equation}
C \psi_L 
= 
\psi_L^* = e^{-i(j_z \phi +\delta)} ||\psi|| = e^{-2i\delta} \psi_R
,
\end{equation}
where $\psi_L$ and $\psi_R$ are Weyl spinors.
We actually see that the parity of $C \psi_L$ and that of $\psi_R$ are the same, and the two are equal up to a global phase $2\delta$ (of course if neglect other complex phases associated with other quantum numbers).
This is the well-known statement that ``the antiparticle of a left-handed particle is a right-handed antiparticle''.
In relativistic physics where all particles are massless, the chirality does not change, so $P \psi$ as well as $C\psi$ are not the same degree of freedom as $\psi$.
This is also consistent with the fact that the gauge interaction, which is chiral, creates a particle and a CP conjugate antiparticle which have both the same spin $j_z=+\frac{1}{2}$, i.e. the same phase modulation $e^{ij_z \phi}$, from the annihilation of a spin-1 ($J_z=+1 = 2j_z$) gauge boson.
The genuine operation to obtain an antiparticle from a particle is therefore the CP transformation.

At the level of the fundamental Lagrangian, all interactions are Lorentz invariant, so their spatial phase modulations all cancel, and the CP transformation properties only depends on the global complex phase $\delta$ of each species of particles.
Each term of the Lagrangian is a product of field operators, so only the relative phases are relevant.
The kinetic term is a combination of a field operator and its complex conjugate, so the global complex CP phase cancels.
For the Dirac fermions, the source of CP violation is therefore the mass (transition $\psi_L \leftrightarrow \psi_R$) or the interactions for which the fermions and their CP conjugates do not appear as pairs.
Since the Lagrangian is Hermitian, complex valued coupling constants and masses generate differences in processes involving particles and antiparticles, which is just the definition of the CP violation.
The essential question is how this asymmetry can be detected.
Let us consider a vector interaction with fermion transition, which is typically contained in nonabelian gauge theory (just imagine a $W$ boson which changes the fermion isospin):
\begin{equation}
{\cal L}_V
=
g_V \bar \psi_1 \gamma^\mu \psi_2 V_\mu
+g_V^* \bar \psi_2 \gamma^\mu \psi_1 V_\mu^\dagger
,
\label{eq:vectorinteraction}
\end{equation}
where the coupling $g_V$ has a complex phase.
Here $V_\mu$ is the vector boson which is in the real representation for simplicity (it is, at least in gauge theory and in hadron physics with vector mesons).
This vector interaction increases the relative phases of the pairs $(\psi_{1L}^*, \psi_{2L} )$ and $(\psi_{1R}^*, \psi_{2R} )$ by arg($g_V$), and decreases exactly by opposite angles those of the Hermitian conjugate pairs.
However, the fermions have chiral symmetry $\psi_{1,2 ; L/R} \to e^{i\theta_{1,2 ; L/R}} \psi_{1,2 ; L/R}$, so that the relative phases may be redefined so as to absorb the complex phase of the coupling $g_V$.
After field redefinition, we may obtain 
\begin{equation}
{\cal L}_V
=
|g_V| \, \bar \psi'_1 \gamma^\mu \psi'_2 V_\mu
+|g_V| \, \bar \psi'_2 \gamma^\mu \psi'_1 V_\mu^\dagger
,
\end{equation}
which is obviously CP symmetric.
We could also have intentionally chosen field redefinitions where $g_V$ has an imaginary part, but this complex phase will either only behave as an overall phase of a scattering amplitude which will disappear at the level of the observables (because the final observables always take the form of a squared absolute value), or will just cancel by appearing by pairs with $g_V^*$, so it has no physical effects.
We therefore need additional interactions to fix the remaining relative phases if we wish to violate CP.
A good candidate is the Yukawa interaction
\begin{equation}
{\cal L}_Y
=
g_Y \bar \psi_1 \psi_2 S
+g_Y^* \bar \psi_2 \psi_1 S^\dagger
,
\label{eq:Yukawainteraction}
\end{equation}
where we again assumed that the scalar field $S$ is a real representation (at least, the SM Higgs boson is).
This interaction fixes the relative phases of the pairs $(\psi_{1L}^*, \psi_{2R} )$ and $(\psi_{1R}^*, \psi_{2L} )$.
A theory with Eqs. (\ref{eq:vectorinteraction}) and (\ref{eq:Yukawainteraction}) has all three relative fermion phases between $\psi_{1L}, \psi_{1R}, \psi_{2L}$, and $\psi_{2R}$ fixed, and one remaining unphysical overall complex phase. 
This last single irrelevant phase cannot be used to absorb simultaneously the complex phases of $g_V$ and $g_Y$, so that at least one imaginary degree of freedom is relevant, violating CP.
Concretely, a loop process which involves the series of creations and annihilations in the order $\psi_{2L} \to \psi_{1L} \to \psi_{2R} \to \psi_{1R} \to \psi_{2L}$ or a scattering with interference between the tree level $\propto g_V$ and the one-loop level $\propto g_V^* g_Y^2$ has a phase shift of arg($2 g_V g_Y^*$) which cannot be erased by field redefinitions.
A theory which involves the vector interaction (\ref{eq:vectorinteraction}) and the Yukawa interaction (\ref{eq:Yukawainteraction}) may therefore violate CP.

A good exercise of finding a physically detectable CP violation is the inspection of the complex phase of the Cabibbo-Kobayashi-Maskawa (CKM) matrix $V_{\rm CKM}$ \cite{Kobayashi:1973fv}, which is a $3\times 3$ flavor transition matrix of the $W$-boson-quark interaction.
This matrix naively has 18 real parameters, but the unitarity $V_{\rm CKM} V_{\rm CKM}^\dagger =\hat 1$ imposes a constraint to each element (i.e. 9 equations), so that we are left with 9 free parameters.
We then have 5 relative phases between 6 quark operators which may absorb 5 parameters.
We finally obtain 4 physical degrees of freedom, which are the three flavor mixing angles and one CP violating phase.
This final complex phase ($\delta_{\rm CKM} \sim 1$ rad) is the sole complex phase which may induce CP violation in the SM of particle physics (the $\theta$-term of QCD will be discussed later).
We note that, as seen for the simple vector (\ref{eq:vectorinteraction}) and Yukawa interactions (\ref{eq:Yukawainteraction}), the appearance of the interaction vertices is not sufficient to violate CP in concrete processes, but they must not appear as pairs of Hermitian conjugates, i.e. a quark must not be created and annihilated by the same (Hermitian conjugate) interaction, including the interfering scattering amplitude.
The CP violation in the SM also occurs through interferences or higher order loops, which require at least four factors of distinct CKM matrix elements \cite{McKellar:1987tf,Buchalla:1995vs,Yamanaka:2015ncb,Yamaguchi:2020eub,Yamaguchi:2020dsy}.
The CP violation by the CKM matrix then suffers from suppression due to flavor mixing angles, and the leading order CP violation cannot avoid the small factor of Jarlskog of $O(10^{-5})$ \cite{Jarlskog:1985ht}.

We just saw that the CP violation of the SM appears after all possible redefinitions of fields.
The extension of the SM proceeds by adding new particles, but these must be integrated out at some high energy scale, and the experimentally detectable effects are most probably due to additional interactions composed of SM fields (often called SM effective field theory, SMEFT \cite{Grzadkowski:2010es,Isidori:2023pyp}).
Since all CP phases have already been absorbed and fixed in the SM, the addition of a new interaction term must violate CP, and the expected ``natural'' size of the complex phase should be of $O(1)$, just like the CKM phase $\delta_{\rm CKM}$.
This is because there are no viable reasons for the new interaction to have the same (or aligned)  complex phases as those of SM interactions, unless there is some physical mechanism behind.
We may then state that the CP violation is an excellent probe of new physics beyond the SM.

\section{Quantification of hadronic CP violation\label{sec:hadronicCPV}}

In this section, we present the current situation of the quantification of hadronic CP violation without flavor violation, which contributes to low energy observables such as the EDM, low energy T-violating neutron optics, and beta decay.
To quantify low energy hadron level processes, we should ideally employ lattice QCD, but calculations are often very difficult, so we frequently use the chiral perturbation theory ($\chi$PT) \cite{Crewther:1979pi,Mereghetti:2010kp,deVries:2015una,deVries:2016jox}.
Phenomena of strong interaction have a typical scale of $\Lambda_{\rm QCD}= O(100)$ MeV, while the QCD Lagrangian has the up and down quark masses $m_{u,d} = O({\rm MeV}) \ll \Lambda_{\rm QCD}$.
We may therefore use $m_{u,d}/\Lambda_{\rm QCD} = O(0.01)$ as an expansion parameter, which appears in processes involving pions.
Fortunately, pions also frequently participate to hadronic CP violation, so we may use $\chi$PT to quantify the hadron level contribution to the low energy observables of interest.

The leading order hadronic CP violating interactions are the nucleon EDM and the CP-odd nuclear force, whose nonrelativistic Hamiltonian is given by
\begin{eqnarray}
H_{\rm NEDM}
&=&
- {\vec d}_N \cdot {\vec E}
,
\label{eq:NEDM}
\\
H_{\rm CPVNN}
&=&
V_{CPV}^{IX} (r)
\Gamma_{IX} (a_{1I} \vec \sigma_1 +a_{2I} \vec \sigma_2 ) \cdot \vec \nabla \frac{e^{-m_X r}}{r}
,
\label{eq:CPVNN}
\end{eqnarray}
where $\vec E$ is the external electric field probing the EDM of the nucleon ($N=p,n$), and $I$ is the isospin and Lorentz structure index, while $X$ is the label of the effectively exchanged meson.
These two interactions give the low energy input of the calculations of nuclear level CP violation \cite{Dobaczewski:2005hz,Yamanaka:2015qfa,Yamanaka:2016itb,Yamanaka:2016umw,Yamanaka:2019vec,Yanase:2020agg,Yanase:2020oos,Froese:2021civ,Yanase:2022ycj,Dutsov:2025kbd,Zhou:2025jhu}.
At an intermediate hadronic scale, we may consider the effective CP violating pionic interaction
\begin{equation}
{\cal L}_{\pi NN}
=
\bar g_{\pi NN}^{(0)} \pi_a \bar N \tau_a N
+
\bar g_{\pi NN}^{(1)} \pi_0 \bar N N
+
\bar g_{\pi NN}^{(2)} ( \pi_a \bar N \tau_a N - 3 \pi_0 \bar N \tau_3 N)
,
\label{eq:piNN}
\end{equation}
which induces Eqs. (\ref{eq:NEDM}) and (\ref{eq:CPVNN}) via radiative processes, and they may be quantified using $\chi$PT up to unknown low energy constants.

The quantification of these low energy constants are the most serious bottleneck of the evaluation, since hadronic CP violation is poorly known, and the experimental data for the flavor diagonal case are even nonexistent.
The most important low energy constants for the hadronic CP violation are given by the matrix elements of the following leading order SMEFT quark-gluon operators:
\begin{eqnarray}
{\cal L}_{\rm qEDM}
&=&
- \frac{i}{2} \sum_q d_q \bar q \sigma_{\mu \nu} F^{\mu \nu} \gamma_5 q
,
\label{eq:qEDM}
\\
{\cal L}_{\rm cEDM}
&=&
- \frac{i}{2} \sum_q d^c_q \bar q \sigma_{\mu \nu} G_a^{\mu \nu} t_a \gamma_5 q
,
\label{eq:cEDM}
\\
{\cal L}_{w}
&=&
\frac{w}{6} f_{abc} \epsilon_{\alpha \beta \gamma \delta} G_a^{\mu \alpha} G_b^{\beta \gamma} G_{\delta , c}^{\ \ \mu}
,
\label{eq:weinbergop}
\\
{\cal L}_{qq'}
&=&
C_{qq'} \bar q i \gamma_5 q \, \bar q' q'
,
\label{eq:CPVqq}
\end{eqnarray}
where $G_a^{\mu \nu}$ is the field strength of the gluon $A^\mu_a$.
These interactions are generated with different Wilson coefficients depending on the model of new physics beyond the SM.
Quantifying the dependences of hadron and nuclear level CP-odd processes on these operators will allow us to determine the elementary level CP violation after comparing with experimental data.
Another important point to clarify is the most dominant source of CP violation at the hadron and nuclear levels, so that this information will help us conceive new experiments.

The above quark-gluon level CP-odd interactions (\ref{eq:qEDM}), (\ref{eq:cEDM}), (\ref{eq:weinbergop}), (\ref{eq:CPVqq}) are generated at the scale of new physics, after integrating out heavy new particles (and also heavy SM particles), but the energy scale defining them (typically TeV) are far from the hadronic scale where low energy constants are matched (around GeV).
The renormalization group evolution is then required to bridge the two scales, and the operators generally mix each other as a result.
This step can be calculated perturbatively \cite{Degrassi:2005zd,Hisano:2012cc,Kley:2021yhn}.
A good example is the Weinberg operator (\ref{eq:weinbergop}) \cite{Weinberg:1989dx}, for which we have 
\begin{equation}
{\cal L}_{w} (\mu = 1\, {\rm TeV}) 
\approx
0.52 {\cal L}_{\rm qEDM} (\mu = 1\, {\rm GeV}) 
-0.59 {\cal L}_{\rm cEDM} (\mu = 1\, {\rm GeV}) 
+0.33 {\cal L}_{w} (\mu = 1\, {\rm GeV}) 
.
\end{equation}
As we can see, the mixing effect is important, which forces us to quantify the matrix elements of all relevant operators.

The simplest CP-odd hadronic process is the quark EDM (\ref{eq:qEDM}) contribution to the nucleon EDM (\ref{eq:NEDM}).
The linear coefficients (called nucleon tensor charge) have been extensively calculated in lattice QCD \cite{FlavourLatticeAveragingGroupFLAG:2024oxs,Yamanaka:2018uud,Hasan:2019noy,Harris:2019bih,Horkel:2020hpi,Park:2021ypf,Tsuji:2022ric,Bali:2023sdi,Wang:2025nsd}, and $\chi$PT is not needed in this case.
The results for the neutron and proton EDMs yield
\begin{eqnarray}
d_n (d_u, d_d)
&\approx &
-0.2 d_u 
+0.8 d_d 
,
\\
d_p (d_u, d_d)
&\approx &
0.8 d_u
-0.2 d_d 
,
\end{eqnarray}
which follow from isospin symmetry, with 10\% of uncertainty.
The renormalization scale of the quark EDMs is $\mu=1$ GeV.
This contribution is not particularly enhanced at the nuclear level.
This can be understood from the suppression of the tensor vertex by the gluonic dressing \cite{Yamanaka:2013zoa}.

\begin{figure}[tbh]
\begin{center}
\includegraphics[width=10cm]{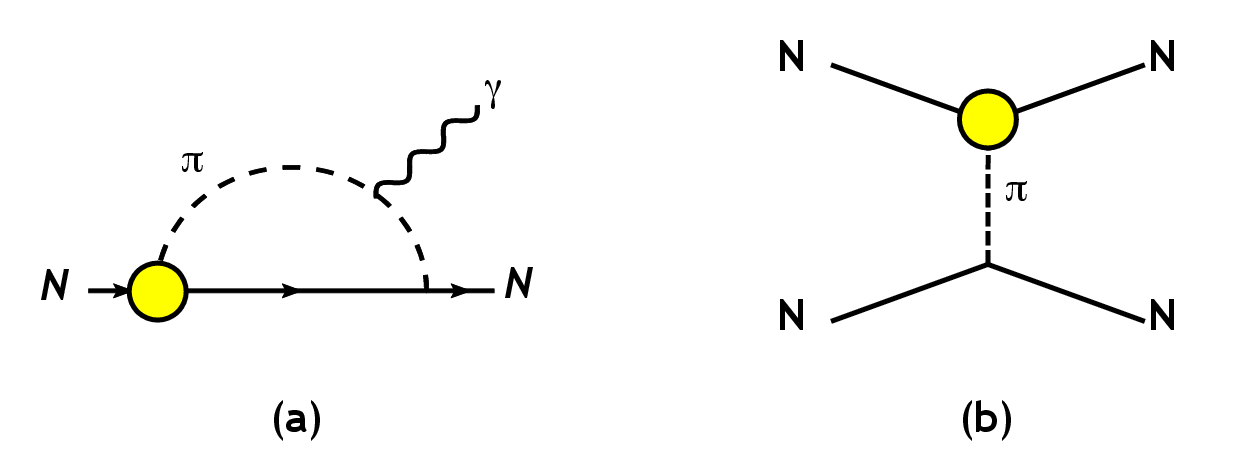}
\caption{
Leading order $\chi$PT contribution to the nucleon EDM (a) and to the one-pion exchange CP-odd nuclear force (b).
The yellow blobs denote the CP-odd pion-nucleon interaction (\ref{eq:piNN}).
}
\label{fig:NEDMCPVNN}
\end{center}
\end{figure}

The nucleon EDM may be generated by the other purely strongly interacting operators (\ref{eq:cEDM}), (\ref{eq:weinbergop}), and (\ref{eq:CPVqq}), but this time $\chi$PT calculations are required.
Given the CP-odd pion-nucleon interaction of Eq. (\ref{eq:piNN}), the leading order $\chi$PT contribution to the nucleon EDM is \cite{Crewther:1979pi,Mereghetti:2010kp,deVries:2015una}
\begin{eqnarray}
d_n 
&\approx &
\tilde d_n
+\frac{e g_A \bar g_{\pi NN}^{(0)}}{8\pi f_\pi} \ln \frac{m_N^2}{m_\pi^2}
+O(m_\pi / m_N)
,
\\
d_p 
&\approx &
\tilde d_p
-\frac{e g_A \bar g_{\pi NN}^{(0)}}{8\pi f_\pi} \ln \frac{m_N^2}{m_\pi^2}
+O(m_\pi / m_N)
,
\end{eqnarray}
where $f_\pi =93$ MeV, $g_A=1.27$, and $\tilde d_p , \tilde d_n$ are the counterterms.
The above $\chi$PT contributions are generated by the loop diagram of Fig. \ref{fig:NEDMCPVNN} (a), but it is not particularly enhanced.
We also see that the isovector CP-odd coupling $\bar g_{\pi NN}^{(1)}$ does not appear at the leading order.

Actually, $\bar g_{\pi NN}^{(1)}$ contributes at the leading order to the CP-odd nuclear force through the diagram shown in Fig. \ref{fig:NEDMCPVNN} (b).
There is a notable enhancement of roughly a factor of ten thanks to the large pion-nucleon sigma term $\sigma_{\pi N} \equiv \frac{m_u+m_d}{2} \langle N | \bar u u + \bar d d  | N \rangle = (45 \pm 15)$ MeV \cite{FlavourLatticeAveragingGroupFLAG:2024oxs,Yamanaka:2018uud,Gupta:2021ahb,Agadjanov:2023efe,Barca:2024hrl,Gubler:2018ctz,Huang:2019crt,Friedman:2019zhc,Huang:2019not,Ikeno:2022foa,Hoferichter:2023ptl}, which is most probably due to the fact that the quark scalar density is increased by the antiquarks of Z-graphs and loops inside nucleons \cite{Yamanaka:2014lva} (at the atomic/molecular level, there is another important effect enhanced by $\sigma_{\pi N}$ which must not be overlooked, the CP-odd electron-nucleon interaction \cite{Yamanaka:2017mef,Yanase:2018qqq}).
We also note that there is a significant discrepancy between the results of $\chi$PT and lattice calculations of $\sigma_{\pi N}$.
The reason of this systematics is presently unknown, but it needs to be fixed in the future in order to quantify the enhancement of the hadronic CP violation.
The isovector CP-odd pion-nucleon coupling depends on the vacuum pion matrix element $\langle \pi_0 | {\cal L}_{CPV} | 0 \rangle$ which, combined with the $\pi -N$ scattering, becomes $\bar g_{\pi NN}^{(1)}$.
By quoting the value of the mixed condensate $\langle 0 | g_s \bar q \sigma_{\mu \nu} F_a^{\mu \nu} t_a q | 0 \rangle \approx 0.8 \langle 0 | \bar q q | 0 \rangle {\rm GeV}^2$ from QCD sum rules analyses \cite{Gubler:2018ctz}, the contribution of the quark chromo-EDM becomes \cite{Sahoo:2023vrn}
\begin{equation}
\bar g_{\pi NN}^{(1)}
\approx
-1 \times 10^{-11} \frac{d_u^c-d_d^c}{10^{-26} {\rm cm}}
,
\end{equation}
where we renormalized $d_u^c$ and $d_d^c$ at 1 GeV.
Let us for instance estimate the EDM of light nuclei.
By multiplying the nuclear level coefficients \cite{Yamanaka:2015qfa,Yamanaka:2016umw,Froese:2021civ}, the EDM of $^3$He is obtained as 
\begin{eqnarray}
d_{\rm He} 
&\approx &
10 \times e (d_u^c-d_d^c)
.
\end{eqnarray}
Since the EDM and the quark chromo-EDM have the same dimension, the coefficient of $O(10)$ shows a clear enhancement.
If we dimensionally compare with other contributions such as the Weinberg operator \cite{Weinberg:1989dx,Demir:2002gg,Haisch:2019bml,Yamanaka:2020kjo,Osamura:2022rak,Yamanaka:2022qlu}, we numerically obtain that $d_u^c, d_d^c$ is the most enhanced.
We also add that the CP-odd 4-quark interaction (\ref{eq:CPVqq}), which has many possible flavor structures, may be enhanced by the pion-nucleon sigma-term if we assume that the factorization approximation works \cite{An:2009zh}.

\section{Resolution of the strong CP problem\label{sec:strongCP}}

The QCD Lagrangian may contain the following CP violating mass dimension-4 $\theta$-term
\begin{equation}
{\cal L}_{\theta}
=
\theta N_f \frac{\alpha_s}{16\pi} \epsilon_{\mu \nu \rho \sigma} G^{\mu \nu}_a G^{\rho \sigma}_a
.
\label{eq:theta-term}
\end{equation}
Here $\theta$ is a free parameter which should be of $O(1)$ according to the naturalness.
However, $\theta$ is known to contribute to the EDM \cite{Crewther:1979pi}, and experimental data are currently putting a strong constraint $|\theta | < 10^{-10}$.
This unnatural fine-tuning is called the ``strong CP problem''.
The most popular way to resolve this problem is to add additional particles or symmetries beyond the SM such as the axion, but there have recently been several proposals to explain it within QCD \cite{Ai:2020ptm,Nakamura:2021meh,Yamanaka:2022vdt,Yamanaka:2022bfj,Ai:2024cnp,Ai:2024vfa,Schierholz:2024var}.

In this section, we review the approach of Refs. \cite{Yamanaka:2022vdt,Yamanaka:2022bfj} which consists of showing the unobservability of the topological charge and the $\theta$-term.
QCD is known to have topologically nontrivial configurations which are labeled by an integer number, and may be probed with the following topological charge operator:
\begin{equation}
\int d^4 x \,
\frac{\alpha_s}{16\pi} \epsilon_{\mu \nu \rho \sigma} G^{\mu \nu}_a G^{\rho \sigma}_a
=
\frac{i g_s \alpha_s}{24\pi}
\int d^3 x \,
f_{abc}
\epsilon_{ijk}
A_{i,a}(x)A_{j,b}(x)A_{k,c}(x)
\Bigg|^{t=+\infty}_{t=-\infty}
=
\Delta n
.
\label{eq:topological_charge}
\end{equation}
Here the integer $\Delta n$ gives the change of the ``winding'' of the gauge field between the infinite past and future.
We remark that the integral of the second equality contains $\epsilon_{ijk}$ which covers all three-dimensional spatial directions.
By noting that the longitudinal component (direction of the gradient or momentum) of the gauge field cannot be observed, we may deduce that Eq. (\ref{eq:topological_charge}) should be as well, since the integral always contains at least one factor of this unphysical mode.
This property is only certified in the perturbation by the BRST symmetry, and one might wonder whether nonperturbative effects might upset the unobservability.
We have to note, however, that the topological charge density operator $\frac{\alpha_s}{16\pi} \epsilon_{\mu \nu \rho \sigma} G^{\mu \nu}_a G^{\rho \sigma}_a$ is generated by the axial anomaly.
It is actually known that the latter is a fermion loop process which is one-loop exact according to Adler-Bardeen theorem \cite{Adler:1969er,Anselmi:2015sqa} and no additional corrections including the nonperturbative ones are allowed.
We may then say that the axial anomaly, and the topological charge which is just its integral, is a perturbative process for which the BRST formalism may be applied.
Consequently, the statement that Eq. (\ref{eq:topological_charge}) which contains the longitudinal gauge mode is unobservable holds (an intuitive way to understand this is that Eq. (\ref{eq:topological_charge}) is integer, but higher order quantum corrections and other nonperturbative effects are non-integer, so they cannot contribute).
We could finally show that the topological charge and the $\theta$-term (\ref{eq:theta-term}) are not observable, so that the EDM does not receive any contribution from $\theta$, which means that the strong CP problem is resolved.
To consolidate the above demonstration, it is desirable to have a robust definition of the nonperturbative effect in quantum field theory, which is still an open question today.

There is actually another very important consequence from the unobservability of the topological sectors.
In the SM, the baryon and lepton numbers ($B+L$) are anomalous \cite{Bardeen:1969md}, and it was believed so far that electroweak topological effects might violate $B+L$ via processes like \cite{tHooft:1976rip}
\begin{equation}
u + d 
\to
\bar d + \bar s + 2\bar c +3\bar t + e^+ + \mu^+ + \tau^+
.
\end{equation}
If true, this is an attractive solution to the matter-antimatter asymmetry generation in the early Universe (sphaleron induced baryogenesis).
However, as discussed above, this scenario is impossible since the topological charge is not relevant in observable processes.
This means that $B+L$ is conserved in the SM and in all simple extensions which do not contain local $B+L$ violating interactions.

\section{Conclusion\label{sec:conclusion}}

In this contribution, we gave an overview of the CP violation, its importance in the search for new physics, the on-going progress of the quantification of the flavor diagonal hadron level CP violation, and the resolution of the strong CP problem.
Since all the complex phase degrees of freedom have already been fixed by the field redefinitions in the SM, the addition of new interactions and fields will certainly contain large CP violation, which is an excellent probe of new physics.
As regard the hadronic CP violation, the quark chromo-EDM contribution to the isovector CP-odd pion-nucleon interaction is the most important effect at the nuclear level.
There is an enhancement due to the quark scalar density of the pion-nucleon sigma-term which is still suffering from large systematics, and its determination is left as an important homework for the quantification of hadron level CP violation.

We also showed that the strong CP problem may be resolved within QCD due to the unobservability of the topological sectors.
This mechanism has also another important consequence, that the sphaleron induced baryogenesis does not occur.
Moreover, we can even conjecture that this unphysicalness of the topological charge also applies to other topological quantities such as the cosmic strings, magnetic monopoles, and domain walls, for the following reason.
A topological object, when observed, will provide the information of infinitely large distance since it spreads throughout the whole space-time (just imagine a cosmic string piercing the Universe).
From the point of view of the quantum uncertainty principle, its detection necessitates an infinitely fine energy-momentum resolution which is of course impossible using any finite size apparatus.
This is a critical quiz for physicists who wish to establish the observability of topological defects.

This work was supported by the RIKEN TRIP initiative (Nuclear transmutation).

\end{document}